# Reconstruction Models for Attractors in the Technical and Economic Processes

E.V. Nikulchev[1]

[1]*Moscow Technological Institute «WTU», Moscow, Russia.*

**ABSTRACT:** *The article discusses building models based on the reconstructed attractors of the time series. Discusses the use of the properties of dynamical chaos, namely to identify the strange attractors structure models. Here is used the group properties of differential equations, which consist in the symmetry of particular solutions. Examples of modeling engineering systems are given.*

***Keywords -*** *Chaotic Dynamics, Symmetry, Identification, Groups of transformations, reconstructed attractors, global reconstruction*

## I.  INTRODUCTION

Research of attractors several times caused a surge of interest of publications. The first burst by the discovery of chaotic dynamics in the Lorenz equations, the second was associated with publications Takens [1] and Packard [2]. This popularity is due to three reasons. Firstly, attractor was simply to build a single observed value from the experimental data. Of course under specified conditions. Secondly, attractor is more informative in a sense than a mathematical model. Kind of Lorenz equations does not say of a dynamic chaos in contrast to the strange attractor. Third, it is just beautiful multidimensional picture. A drop in popularity of methods can be explained by the fact that engineering practice a constructive method is required. That is, the picture must be associated with any particular tool set. Equation obtained either bulky or ugly. Specialists accustomed to dynamic chaos that chaos is observed in accurate equations of small dimension. To solve the problem requires a huge number of assumptions and hypotheses, with the number of "correct" decisions too much. On the points you can build any number of equations. The situation is complicated by the fact that the authors of simulation methods of chaotic systems are not considered data from the sensors real processes and simulated models.

## II.  NEW CONSTRUCTIVE APPROACH

The method of the global reconstruction of a dynamic system of equations for its one dimensional realization was proposed in [3, 4]. The algorithm is as follows. One-dimensional realization of the process in a system, which is considered a "black box" recovered phase portrait on the Takens theorem, topologically equivalent to the attractor of the original system. According to a priori given equations, is the method of least squares a set of unknown coefficients. There are some modifications of this approach. For example, in [5] to reconstruction dynamic equations on the experimental time series with a broadband continuous spectrum use additional information about the dynamic and statistical properties of the original system contained in the implementation. In obtaining equation takes into account the values of Lyapunov exponents and the probability density, calculated from the original time series. However, the resulting evolution equations have a very cumbersome, inconvenient to use. In [5] used the hidden variables to write model equation. In [6] describes a method for synchronizing the model with the original data. Now there is publications, developing and constantly improving the proposed the method [7, 8, 9 etc.]. The review [10] provides a detailed list of works.

However, all these methods are inherently not use inherently chaotic properties. They are just trying to recover equation based on time series.

Proposed new method for finding the structures of equations based on the geometric properties of differential equations [11]. Back in the late 19th century Sophus Lie was proved that the transformations that particular solutions of a differential equation form a group, where the identity element of the group is the identity transformation. If one knows a group of transformations that can be one particular solution to restore the general solution of the differential equation. Unfortunately, all the conversion cannot be known, but they can search.

The essence of this method is. Consider some parts of the attractor and try to establish the fact of whether there is, or that the transformation in the solution of the differential equation. A solution - this is the phase portrait of which we have restored Takens theorem.

It is also certainly a daunting task, but solvable. In strange attractors is a symmetry shift, rotation, stretching and compression. This is used in practical applications to find transformations. For the class of affine systems availability symmetry determines the form of the nonlinear component.

### III. FORMAL STATEMENT

Let us consider a model system in the form of a system of differential equations of the form

$$\frac{dx}{dt} = Ax(t) + \Phi(x,t), \qquad (1)$$
$$y = Cx,$$

where $y$ — observed value, $x$ — $n$-dimensional state vector of the system, $t$ — time, $A$ — matrix of nxn, $\Phi(x,t)$ — is a $\mathbf{C}^r$-smooth function, whose structure is determined by the symmetry transformation.

Note that lead to the canonical form of the system in the traditional form [see e.g. 12] can be simple transformations.

For a discrete system, the local center manifold is determined by the system:

$$x(k+1) = Ax(k) + \Phi(x,k),$$
$$y(k) = Cx(k), \qquad (2)$$

where $k$ — discrete time, $\Phi(x,t)$ — is a $\mathbf{C}^r$-smooth function defined on the basis of symmetry groups constructed on the reconstructed attractor.

For the reconstruction of a nonlinear system in the form (1) (2) proposed the allocation of local regions of phase trajectories and the construction of finite transformations one area to another. That is, the construction of the symmetry group of phase trajectories, which is characterized by the transformation of particular solutions at intervals. The resulting transformations determine the structure of the desired evolution equations.

### IV. IMPLEMENTATION IN MATLAB

The implementation consists of two parts. To find sites and identify transformations developed genetic algorithm [13].

The presence of symmetries in accordance with the formula Hausdorff- Lee gives a view nonlinear form. Well, in general, understood that it will either sine or exponential, depending on what kind of transformation the most prevalent.

Actually parametric identification is carried out several specific ways. There are standard dynamic model in state space. In this model, instead of the control is used on the basis of symmetry found nonlinear function. Coefficients are themselves considered to be unknown and subject to calculation. It uses just the same kind of pattern as well as the independence of the state space and offices. Thus, identifying the input pattern time series fed source, the control signal is found by the function corresponding to the revealed transformation.

Identification Toolbox [14] is used due to its powerful tools associated with the ability to filter, select models with delay and forecasting, etc. Incidentally dimension found entrance solutions coincides with the dimension of the system obtained in the reconstruction of the attractor.

However, we must recognize that the instability of the method of least squares to find the coefficients of the system equations. Perhaps for some practical problems need to use robust methods, then ensure stability of the solution at a predetermined interval.

### V. EXAMPLES

Method was published in Russian and some scientists began to use it in their applications for identification of dynamical systems - in geology and metallurgy, in the financial sector [14, 15]. Give here examples of their practical applications.

*Test Example. Rösler system.*

Let observable parameters generated by the system:

$$\dot{x}_1 = -(x_2 + x_3),$$
$$\dot{x}_2 = x_1 + 0.2x_2,$$
$$\dot{x}_3 = 0.2 + x_3(x_1 - 5.7),$$

Simulation result is shown in Fig. 1, phase trajectories of the resulting model in Fig. 2.

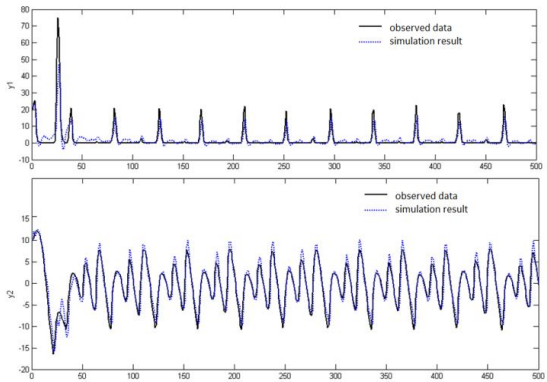

Fig. 1. Comparison of the dynamics of the test series and the constructed model

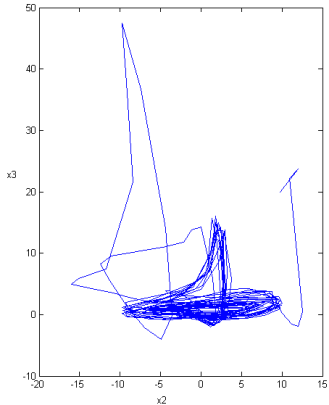

Fig. 2. Phase trajectories of the resulting model

*Example 2.* **Aluminum cooling process.**

Here y1 - the cooling rate of the alloy, y2 - coolant flow. Fig. 3 shows the attractor of the system. Results of the comparison of the dynamics of the simulated process with real data are shown in Fig. 4.

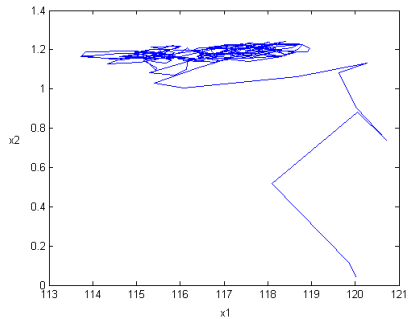

Рис.3. Phase trajectories of the resulting model of cooling process

Reconstructed equation has the form (2) with the corresponding coefficients:

$$A = \begin{bmatrix} 0,7987 & 0,5871 & -0,1104 & -0,0795 & -0,0100 & 0,0100 & -0,0062 & 0,0283 & -0,0383 \\ -0,5605 & 0,6582 & -0,2665 & -0,3695 & 0,0204 & -0,3091 & -0,1216 & 0,0319 & 0,0308 \\ -0,1076 & 0,2538 & 0,9248 & -0,2668 & -0,0639 & -0,3874 & -0,2682 & -0,3668 & 0,6859 \\ -0,1254 & 0,2899 & -0,0054 & 0,7027 & -0,0487 & -0,7902 & 0,0226 & -0,1279 & 0,6137 \\ -0,0337 & 0,0967 & 0,2727 & -0,0898 & 0,7439 & -0,6723 & -0,3121 & -0,0916 & 0,6098 \\ -0,0186 & 0,0371 & -0,2327 & 0,2056 & 0,2983 & 0,4822 & 0,0862 & -0,3811 & -0,2154 \\ -0,0062 & 0,0477 & 0,2632 & -0,0479 & -0,0213 & -0,7668 & 0,0473 & -0,2913 & -0,1507 \\ 0,0110 & 0,0150 & 0,1896 & 0,1428 & -0,3417 & -0,0715 & -0,3949 & 0,5512 & -0,0141 \\ 0,0537 & -0,0289 & -0,5188 & 0,0982 & -0,8413 & 1,5644 & 0,7244 & -0,1601 & -0,6049 \end{bmatrix};$$

$$\Phi = \begin{bmatrix} -3,9240 & 0,7319 \\ 1,7215 & -1,3336 \\ 2,0934 & -1,6825 \\ 2,4342 & -0,8467 \\ -0,2892 & -0,8989 \\ 2,7220 & 1,4839 \\ 0,4833 & 1,0770 \\ 1,0878 & -0,0228 \\ 2,8889 & 0,5654 \end{bmatrix} \begin{bmatrix} t^2 - 2t - 0,93; \\ \sin(t-10); \end{bmatrix};$$

$$C = \begin{bmatrix} 0,1857 & 0,0442 & -0,0082 & -0,0066 & -0,0005 & -0,0012 & -0,0010 & 0,0019 & -0,0022 \\ 0,6124 & 0,5277 & -8,6861 & 9,1780 & -1,1570 & -2,1478 & -2,6497 & 3,0497 & -0,3083 \end{bmatrix}.$$

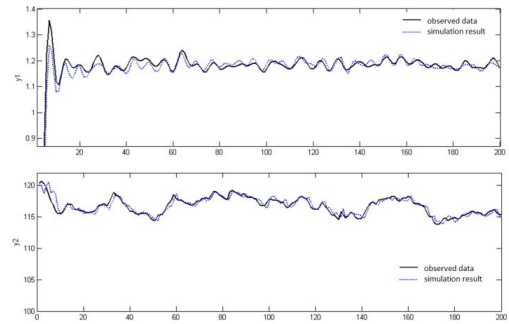

Fig. 4. Comparison of experimental data and modeling

*Example 3.* **The heating of the viscous fluid.**

Application of the equation gives the corresponding values:

$$A = \begin{pmatrix} 0,998 & 0,006 & -0,009 & 0,007 & -0,001 & -0,007 \\ 0,008 & 0,913 & 0,238 & 0,142 & 0,091 & -0,215 \\ 0,006 & 0,132 & 0,091 & -0,928 & -0,459 & -0,038 \\ -8,12 \cdot 10^{-5} & 0,008 & -0,177 & 0,356 & -0,956 & 0,097 \\ -0,001 & -0,06 & 0,803 & 0,203 & -0,145 & 0,139 \\ 0,002 & -0,006 & 0,006 & -0,003 & 0,010 & 0,661 \end{pmatrix}$$

$$\Phi = \exp(10t) \left( -7,24 \cdot 10^{-4}; -0,087; 0,209; -0,074; -0,440; -0,061 \right)^T$$

$$C = \begin{pmatrix} 129,8 & -144,2 & 18,43 & -23,21 & -3,233 & -4,911 \\ 582,1 & 4,995 & 2,230 & -1,596 & 2,587 & 4,100 \end{pmatrix}$$

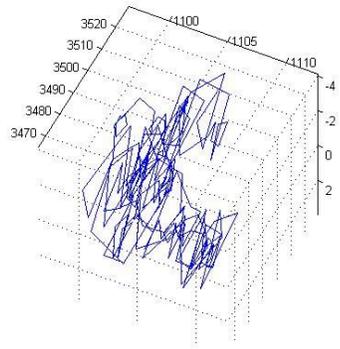

Fig. 5. Reconstructed phase portrait

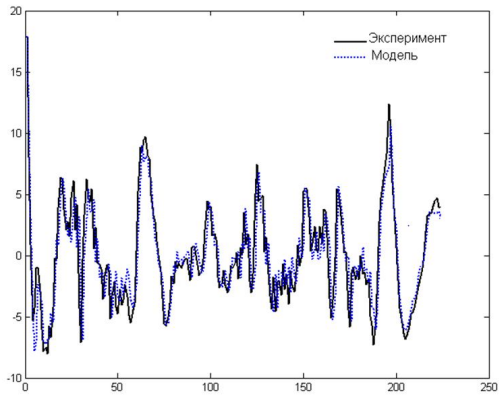

Fig. 6. Comparison of experimental data and modeling

*Example 4.* **Financial row.**

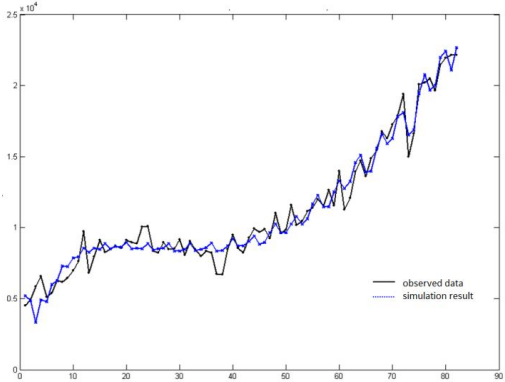

Fig. 7. Comparison of experimental data and modeling

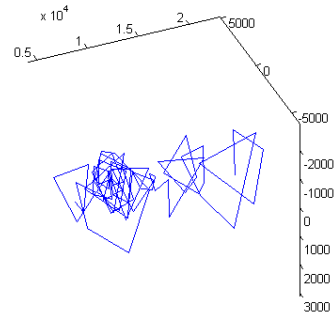

Fig. 8. Phase trajectories of the resulting model of cooling process

*Example 5. Simulation of traffic.*

It is noteworthy that a number of researchers have used the chaotic model for building traffic models [12, 18]. The presence of the symmetry of the system determined the structure of the form (2). Parameter identification system using the method of least squares, gives the following result:

$$A = \begin{bmatrix} 0.9413 & -0.1805 & 0.1164 & -0.0295 \\ -0.0545 & 0.8226 & 0.1622 & 0.1056 \\ 0.0014 & -0.0105 & -0.4455 & 0.8471 \\ -0.0062 & 0.0341 & -0.8860 & -0.5404 \end{bmatrix},$$

$$\Phi = \begin{bmatrix} 0.0399 \\ 0.0463 \\ -0.4848 \\ -0.1851 \end{bmatrix} \left( \exp(t^{0.0001}) \sin(t^{0.4}) \right),$$

$$C = 10^4 \begin{bmatrix} 2.1037 & -0.0124 & 0.1202 & -0.0302 \end{bmatrix}.$$

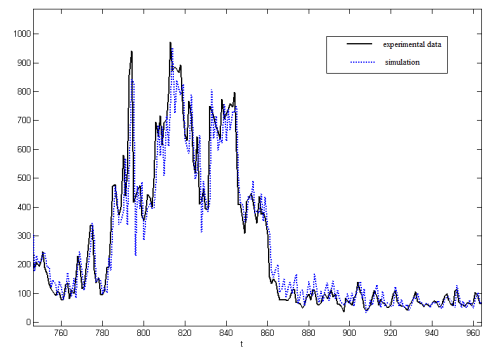

Fig. 9. Comparison of experimental data and modeling

These examples identify determine the accuracy of the proposed method and its effectiveness for modeling different systems.